\documentstyle[mprocl,epsfig]{article}

\def\Journal#1#2#3#4{{#1} {\bf #2}, #3 (#4)}

\def\NPB{{\em Nucl. Phys.} B}
\def\PLB{{\em Phys. Lett.}  B}
\def\PR{\em Phys. Reports}
\def\PRL{\em Phys. Rev. Lett.}
\def\PRD{{\em Phys. Rev.} D}
\def\ZPC{{\em Z. Phys.} C}
\def\pramana{\em Pramana}

\def\be{\begin{equation}}
\def\ee{\end{equation}}
\def\bea{\begin{eqnarray}}
\def\eea{\end{eqnarray}}
\def\to{\rightarrow}

\def\ben{\begin{subequations}}
\def\een{\end{subequations}}
\def\eplem{\mbox{$e^+e^-$}}
\def\gamgam{\mbox{$\gamma \gamma$}}
\def\mh{\mbox{$m_h$}}
\def\mhzer{\mbox{$m_{h_0}$}}
\def\hzer{\mbox{${h_0}$}}
\def\mHzer{\mbox{$m_{H_0}$}}
\def\mt{\mbox{$m_t$}}
\def\mz{\mbox{$m_Z$}}
\def\ma{\mbox{$m_A$}}
\def\gluglu{\mbox{gg}}
\def\phione{\mbox{$\Phi_1$}}
\def\phitwo{\mbox{$\Phi_2$}}
\def\mathv{\mbox{$\mathit v$}}
\def\lsim{\:\raisebox{-0.5ex}{$\stackrel{\textstyle<}{\sim}$}\:}
\def\gsim{\:\raisebox{-0.5ex}{$\stackrel{\textstyle>}{\sim}$}\:}

\begin{document}
\begin{flushright}
CTS-IISc/5/99\\
hep-ph/9904381\\
\end{flushright}
\title{ Predictions for Higgs and SUSY Higgs properties and
their signatures at the Hadron Colliders.\footnote{Invited talk 
at the 13th Topical Conference on Hadron Collider Physics,
Jan.14-20, 1999, Bombay, India.}}
\author{R.M. Godbole}
\address{ Center for Theoretical Studies, Indian Institute of Science, 
Bangalore, India.\\E-mail: rohini@cts.iisc.ernet.in}
\maketitle\abstracts{ In this talk I shall present a discussion of 
the theoretical bounds on the mass of the Higgs in the Standard Model 
(SM) as well as in the Minimal Supersymmetric Standard Model (MSSM). Then
I will point out a few facts about the couplings of scalars that
are relevant for its search at hadronic colliders. After that I discuss
the search possibilities at the Tevatron and the LHC, paying special
attention to the issue of how well one can establish the quantum 
numbers and the couplings of the Higgs, when (if) it is discovered.}
\section {Introduction}
I would concentrate here on the theoretical bounds on Higgs mass
both in the SM and the MSSM as well as on the theoretical information
about its couplings which are relevant for the Higgs search at the 
Tevatron and the LHC. This would be followed by a discussion of the search
prospects for the Higgs at  both these colliders.  I will also
address the issue of the feasibility of establishing the quantum numbers
and the couplings of a spin zero particle when it is discovered at these
colliders.  This would be necessary to establish it as {\bf the } Higgs
boson which arises from the Higgs mechanism of Spontaneous Symmetry
Breakdown (SSB). We will see that to achieve the latter,
search at the hadron colliders needs to be complimented by that at 
a high energy \eplem\ collider 
(the next linear collider NLC)~\cite{physrep}.

\section{Higgs Couplings and masses: Theoretical predictions}
\subsection{Predictions in the SM:}\label{twoone} 
In the SM the existence of  Higgs boson is necessary to 
bring about the SSB which gives masses to  the fermions and the  
gauge bosons, still keeping the theory renormalisable.  For the 
SSB to happen, the $\mathrm{(mass)}^2$ term for the complex scalar doublet
$\Phi$ has to be negative, i.e.  the potential $V(\Phi)$ is 
\be
V(\Phi) = {\frac {\lambda} {4!}} (\Phi^\dag \Phi)^2 - \mu^2 \Phi^\dag \Phi
\label{e1p}
\ee
with $\mu^2$ positive. After the SSB, out of the four scalar fields
which comprise $\Phi$, we are left with only the physical scalar $h$
with a mass 
\be
m_h^2 = \lambda \mathit{v}^2
\label{e1}
\ee
Further, the tree level couplings of the Higgs boson $h$ to the SM
fermions and the gauge bosons are uniquely determined  
and are proportional to their masses. The coupling of a Higgs
to a pair of gluons/photons does not exist at the tree level,
but is induced at one loop level by the diagrams shown in 
\begin{figure}[htb]
\begin{center}
\mbox{\epsfig{file=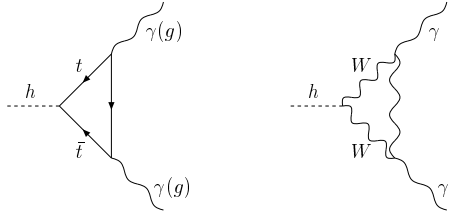,height=30mm}}
\end{center}
\caption{Loop diagrams responsible for $h \rightarrow \gamgam (\gluglu)$.
\label{figure1}}
\end{figure}
Fig.~\ref{figure1}.
As with the other couplings, this coupling is also completely calculable 
given the particle content of the SM, to a given order in the strong
and electromagnetic coupling~\cite{spira}. The $h$gg coupling is
dominated by the top quark contribution in the loop diagram whereas
the $h$\gamgam\ coupling receives dominant contribution from both,
the top loop as well as the $W$ loop. These channels have appreciable
branching ratios, albeit very small, only for $m_h \lsim 2 m_W$. Recall
also that the precision measurements at the $Z$ indicate $\mh \lsim 300$ 
GeV~\cite{LEPCH}.  It should be noted that the QCD corrections for 
$\Gamma (h \rightarrow \gluglu)$ are significant $(\sim 65 \%)$~\cite{spira}. 

In the intermediate mass range (i.e. $m_h \leq 140$ GeV) the total width
of the Higgs is  $\lsim 10 $ MeV, dominant decay is into a $b \bar b$ final
state (e.g. for $m_h = 120 $ GeV $\Gamma (h \rightarrow b \bar b)
\simeq 68 \% )$; on the other hand the branching ratio into a \gamgam\ final
state is about one part per mille. The total width is $\sim 1 $ GeV around
$m_h \sim 300 $ GeV and rises very fast after that, reaching $\Gamma_h  \sim 
m_h$  around $m_h \gsim $ 500 GeV. Calculations of various branching ratios, 
including higher order QCD effects are available~\cite{spira}.

While the various couplings and hence the branching fractions of the Higgs
are well determined once \mh\  and  various  other parameters in the 
SM such as $m_t, \alpha_s$ etc. are specified, \mh\ itself is completely
undetermined in the SM.  However, as seen from Eq.~\ref{e1}, it is linearly
related to the self coupling of the scalar field. Eventhough, the theory has
nothing to say about the Higgs mass {\it per se} the behaviour of the self
coupling $\lambda$ is determined by field theory. This then puts bounds on 
\mh . These bounds can be understood as follows. The self coupling receives 
radiative corrections from the diagrams indicated in 
\begin{figure}[htb]
\begin{center}
\psfig{figure=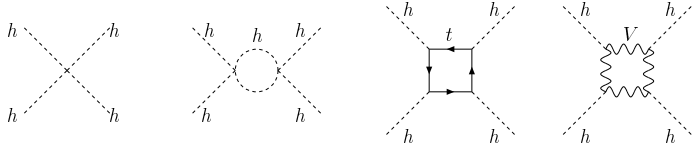,height=25mm}
\end{center}
\caption{Diagrams responsible for radiative corrections to self coupling.
\label{figure2}}
\end{figure}
Fig.~\ref{figure2}. The contributions from the diagrams involving the
scalar and the gauge boson loops on one hand and the fermion loops on
the other, are opposite in sign. The requirement that the self coupling 
$\lambda$ stay positive, i.e.
the vacuum remain stable under radiative corrections, puts a lower bound on 
\mh\ for a given value of $m_t$~\cite{ltop}. This bound however, 
depends on the $h t \bar t$ coupling and hence can be evaded in models 
with more than one Higgs doublets.
\mh\ is also bounded from above by considerations of triviality. This can be
understood by considering only the contributions of the scalar loops, for 
simplicity, to the radiative corrections  to $\lambda$ . It can be shown that
the self coupling then satisfies, 
\be
\frac{d \lambda(t)}{dt} = \frac{3}{4 \pi^2} \lambda^2(t)
\label{e2}
\ee
with $t = \ln(\Lambda /{\mathit v})$; where $\Lambda$ is the momentum scale at
which the coupling $\lambda$ is evaluated. This equation needs a boundary 
condition to solve it, which is chosen as 
\be
\lambda = \lambda(\Lambda={\mathit v}) = \lambda(1) = \sqrt{2} G_F m_h^2.
\label{e3}
\ee
Then Eq.~\ref{e2} above can be solved to give
\be
\lambda (t) = \frac{\lambda}{1 - {3 \lambda t}/{4 \pi^2}}
\label{e4}
\ee
This shows that $\lambda(t)$ will diverge at high scales. If we demand
that the Landau pole, where $\lambda (t)$ will blow up, lies above a scale
$\Lambda$, we get 
\be
m_h \lsim \frac{893}{\sqrt{\ln (\Lambda/\mathv)}} {\mathrm GeV}.
\label{e5}
\ee
Hence the requirement that the theory be valid at large $\Lambda$ and 
yet be nontrivial  at a scale \mathv\ , puts an upper limit on 
$\lambda (\mathv)$ and hence on \mh\ due to the identification in 
Eq.~\ref{e1}. The above analysis, of course, has  to be improved using
the renormalisation group equation~\cite{sher,lindner}. As $\lambda$  becomes 
large, clearly perturbative methods used above must fail. In the region of 
large $\lambda$  the analysis has been done using lattice 
theory~\cite{AH}.

The lower bound on \mh\ implied by the vacuum stability arguments~\cite{ltop}
and the upper bound implied by the triviality 
considerations~\cite{sher,lindner}, depend on the value 
of \mt\ and the uncertainties in the nonperturbative dynamics respectively. 
The resulting bands, taking into account these theoretical uncertainties,
are shown in 
\begin{figure}[htb]
\begin{center}
\vspace{-0.3cm}
\mbox{\epsfig{file=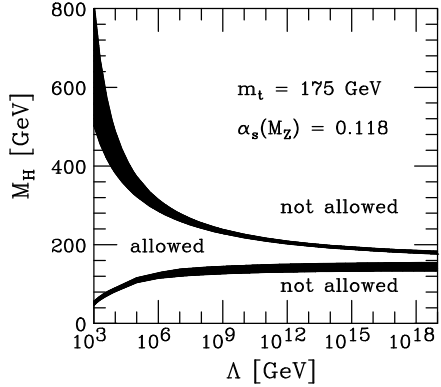,height=60mm}}
\vspace {- 0.7 cm}
\caption{Theoretical bounds on \mh\ in the SM.
\label{figure3}}
\end{center}
\end{figure}
Fig.~\ref{figure3}~\cite{hambye}.

The results can be summarised as follows:
\begin{enumerate}
\item If we demand that the Landau pole lies above $10^{15} (10^{18})$ GeV
and there exists no new physics other than the SM upto that scale,  then
$\mh < 190 (130)$ GeV.
\item If we assume that the SM is an effective theory only upto 1 TeV, i.e.,
there is some new physics at that scale then $m_h \lsim 800 $ GeV. As an aside
let us also mention here that this upper limit of 800 GeV is of the same order
as the limit obtained by requiring that $WW \rightarrow WW (ZZ \to ZZ) $ 
amplitude satisfies perturbative unitarity~\cite{llsmith,thacker}. 
\end{enumerate} 

\subsection {Masses  and couplings in the MSSM}
In the MSSM there exist two complex Higgs doublets $\Phi_1, \Phi_2$ with 
hypercharge $Y = \pm$ respectively. As a result there are five physical 
degrees of freedom left after the spontaneous symmetry breakdown. The
MSSM thus contains, in all, five scalars: three neutrals out of which 
two are CP even states denoted by $h_0, H_0$ and one is a CP odd state denoted 
by A and a pair of charged Higgs bosons $H^\pm$.  Thus the scalar sector of the
MSSM is much richer than in the SM.  $h_0$ denotes the lighter of the 
two CP even neutral scalars.

The most general scalar potential for a two Higgs doublet  model contains 
a large number of free parameters, essentially analogues of the self coupling
$\lambda$ and $\mu^2$ term in the case of the SM. However, supersymmetry 
either fixes all these self couplings in terms of the gauge couplings or 
requires them to vanish. Hence the scalar potential for the MSSM is
\bea
V &=& m_{11}^2 \phione^\dag \phione  +  m^2_{22} \phitwo^\dag \phi2 
      - \left[ m^2_{12} \phione^\dag \phitwo + {\mathrm h.c.} \right]\nonumber\\
       &&+\frac {1}{8} (g^2 + g'^2 ) \left[ (\phione^\dag \phione)^2 
       + (\phitwo^\dag \phitwo)^2 \right]\nonumber \\ 
      &&+\frac{1}{4} (g^2 - g'^2)(\phitwo^\dag \phitwo) (\phione^\dag \phione) 
       - \frac{1}{2} g^2 (\phione^\dag \phitwo)(\phitwo^\dag \phione),
\label{e6}
\eea
where $g,g'$ are the $SU(2), U(1)$ coupling constants respectively. 
After the SSB where the neutral members of the two doublets acquire 
vacuum expectation values $\mathv_1/\sqrt{2}, \mathv_2/\sqrt{2}$ respectively, 
the resulting masses of the five physical scalars that follow from the 
abovementioned potential, can be written in terms of two parameters which
can be chosen to be $m_A, \tan \beta = \mathv_1 / \mathv_2$ or
equivalently $m_{H^\pm}, \tan \beta$.  At the tree level the five scalar
masses satisfy the following inequalities:
\be
\mhzer \leq m_Z,\;\;\; \mHzer > m_Z,\;\;\; m_{H^\pm} > m_W,\;\;\; \mhzer < \mHzer, m_{H^\pm} .
\label{e7}
\ee
In the decoupling limit~\cite{haber1} ($m_A \to \infty $) one finds that, 
independent of $\tan \beta$,  all the four heavy scalars become degenerate and
infinitely heavy and the mass of the lightest scalar approaches the upper bound.
In this limit the couplings of the $h_0$ to matter fermions and the gauge 
bosons approach those of the SM higgs $h$. The interesting thing to note here
is that all the masses $m_{H^\pm}, m_A, \mhzer$ can become large without some 
self coupling becoming strong, unlike the case of the SM. 

If we denote by $\alpha$ the  mixing in the two CP even neutral fields to
yield the mass eigenstates $h_0,H_0$, then the couplings of all the 
three neutral scalars, with the fermions and gauge bosons
are given in Table~\ref{table1}.
\begin{table}[htb]
\begin{center}
\caption{Couplings of the three neutrals with fermions and gauge bosons
\label{table1}}
\vspace{0.2cm}
\footnotesize  
\begin{tabular}{|c|c|c|c|c|}
\hline
&&&&\\
& {$h$} & {$h_0$} &{$H_0$} & A\\ 
&&&&\\
\hline
&&&&\\
{$b \bar b$}&{${g m_b}/{2m_W}$}&{$\sin \alpha /\cos\beta$}
&{$\cos \alpha /\sin \beta$}&{$\tan \beta$}\\ 
&&{$\rightarrow 1$}&{$\tan \beta$}&{$\tan \beta$}\\
\hline
&&&&\\
{$t \bar t$}&{${g m_t}/{2m_W}$}&{$\cos \alpha /\sin \beta$}
&{$\sin \alpha /\cos \beta$}&{$\cot \beta$}\\ 
&&{$\rightarrow 1$}&{$\cot \beta$}&{$\cot \beta$}\\
\hline
&&&&\\
{$VV$}&{$g m_V$}&{$\sin (\beta - \alpha)$} 
&{$\cos (\beta - \alpha)$}&0\\ 
&&{$\rightarrow 1$}&0&0\\
\hline
\end{tabular}
\end{center}
\end{table}   
For the  MSSM scalars only the additional factors relative to the
SM case are written down. 
These couplings in the {\it decoupling limit} reduce to the ones denoted in
the second line in each case. We notice that  in the decoupling limit, 
where the
upper limit on the mass of the lightest scalar is saturated, the couplings of 
the lightest scalar approach those for the SM higgs $h$. It should also be 
noted that the CP odd scalar A does not couple to a pair of gauge bosons
at the tree level.

The inequalities of Eq.~\ref{e7} get affected by the radiative corrections 
to the Higgs masses. The dominant corrections arise from loops involving
top (t) and its scalar partner stop ($\tilde t$)  due to the large Yukawa 
coupling of the top quark. There are many different methods, involving
different methods of approximations~\cite{haber2}, to calculate these 
corrections. These depend on \mt\ as well as  the masses of and the mixing
between $\tilde t_L$, $\tilde t_R$ (the superpartners of $t_L$ and $t_R$). 
To a good approximation, the radiatively corrected upper bound on the
mass of the lightest CP neutral scalar (\mhzer), can be written as
\be
m^2_{h_0} < m_Z^2 \cos^2 2\beta + \epsilon + \epsilon_{\mathrm mix},
\label{e8}
\ee
where 
\bea  
\epsilon &=& \frac{3 g^2 m_t^4}{8 \pi^2 m^2_W}  
\ln \left(\frac{m^2_{\tilde t}}{m^2_t}\right) \nonumber \\
\epsilon_{\mathrm mix}&=& 
\frac{3 g^2 m_t^4}{8 \pi^2 m^2_W} \frac {A^2_t}{m^2_{\tilde t}} 
\left(1 - \frac{A^2_t}{12 m^2_{\tilde t}}\right),
\label{e9}
\eea
with $A_t$ being the coefficient of the trilinear, supersymmetry breaking term,
and $m_{\tilde t}$ being the common mass of the $\tilde t_L, \tilde t_R$. The 
second term $\epsilon_{\mathrm mix}$, even though dependent on $A_t$, 
can be shown to  be  bounded by ${9 g^2 m^4_t}/{8 \pi^2 m_W^4}$. Thus 
\mhzer\ is still bounded eventhough the bound on \mhzer\ of Eq.~\ref{e7} is 
changed by radiative corrections. Also note that the corrections 
will vanish in the limit of exact supersymmetry. The limits on the 
radiatively corrected scalar masses for the case of maximal mixing in 
\begin{figure}[htb]
\vspace{1cm}
\begin{center}
\mbox{\epsfig{file=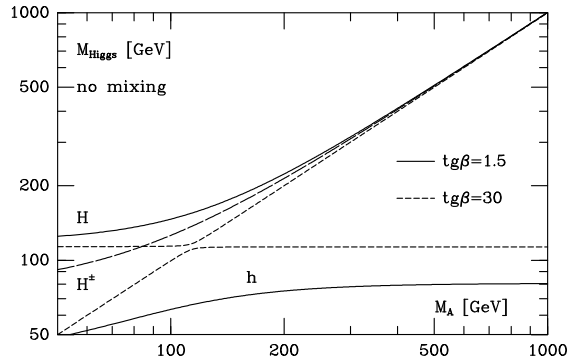,height=45mm}}
\vspace{-0.5cm}
\end{center}
\caption{Bounds on the masses of the scalars in the MSSM \label{figure4}}
\end{figure}
the stop sector are shown in Fig.~\ref{figure4}~\cite{abdel}.
It can be seen from the results shown in the figure
that the mass of the lightest scalar in MSSM  is bounded by $\sim 130 $ GeV 
even after it is radiatively corrected. This bound does get modified 
in the NMSSM~\cite{kane,quiros,pandita}. Again, it has been shown 
that for all reasonable values of the model parameters, \mhzer\ 
is bounded by $\sim 150$ GeV. 
From the figure, it might seem that the current LEP bound on the mass of
the \mhzer , \mHzer and $m_A$ rule out the low $\tan \beta$ values for the MSSM. 
However, it has been shown~\cite{dpman} that a rather minor extension of the 
MSSM, can help avoid this conclusion.  
The couplings of Table~\ref{table1} do get modified to some extent by the 
radiative corrections, but the general features discussed above remain
unchanged.

In discussing the search strategies and propspects of the MSSM scalars one has 
to remember the following important facts:
\begin{enumerate}
\item Due to the reduction of the $h_0 W W$ coupling as indicated in 
Table~\ref{table1}, the $h_0 \gamma \gamma$ coupling is suppressed as compared 
to the corresponding SM case. Of course one also has to include the
contribution of the charged sparticles in the loop~\cite{abdel2}.
Due to the upper limit on \mhzer\ the
decay mode into $WW (VV)$ pair is not possible for $h_0$, due to kinematic
reasons. On the other hand, for $H_0$ the  suppression of the coupling to $VV$ 
as seen from Table~\ref{table1}, makes the decay less probable as compared to
the SM case. As a result, the MSSM scalars are expected to be much narrower
resonances as compared to the SM case. For example, the  maximum width of 
$h_0$ is less than few MeV, for reasonable values of $\tan \beta$ 
and even for the heavier scalars $H_0, A$,  the width is not more than few tens
of GeV even for masses as high as 500 GeV. 

\item $h_0$ is much narrower than the SM Higgs.
However, over a wide range of $\tan \beta, m_A$ values, the $h_0$
has dominant decay modes into Supersymmetric particles. The most interesting
ones are those involving the lightest neutralinos, which will  essentially
give `invisible' decay modes to the $h_0, H_0$ and $A$ ~\cite{abdel1}.

\item On the whole for the MSSM scalars the decay modes into 
fermion-antifermion pair are the dominant ones due to the point 
(1) above as well as the fact that the CP odd scalar A does not have 
any tree level couplings to $VV$. Hence,
looking for the $\tau^+ \tau^-$ and $\bar b b$ final state becomes very
important for the  search of the MSSM scalars.
\end{enumerate}
It is clear from the above that the phenomenology of the  MSSM scalars is
much richer and more complicated than the SM case. Again, calculation 
of various decay widths including the higher order corrections has been 
done~\cite{spira,abdel2}.

\section{Production and search of Higgs at Colliders}
In this section we will begin by discussing the search possibilities 
for the SM Higgs $h$. As is clear from the discussions in the earlier 
section of the couplings 
of the Higgs, the most efficient way of producing the Higgs scalar at any 
Collider is through its coupling to gauge bosons or to a heavy
fermion-antifermion ($t \bar t$) pair. At the hadronic colliders 
the following processes can contribute  to the production:
\bea
\label{e10}
 gg  &\to  & h \\
\label{e11}
q \bar q' &\to & h W\\
\label{e12}
q \bar q &\to&  h Z \\
\label{e13}
 qq &\to&h q q \\
\label{e14}
gg, q \bar q &\to& h t \bar t, h b \bar b.
\eea
At the Tevatron energies ( for Run-II/TeV33) the most efficient processes 
are those of  Eqs.~\ref{e10},\ref{e11}. The potential of direct Higgs 
search at the Tevatron is discussed elsewhere in the proceedings~\cite{paulg}.
The associated $W/Z$ can make it possible to use the dominant $b \bar b$
decay mode. However, the energy of the Tevatron is just too small to give
appreciable production cross-section, except  for the values of  \mh\
close to its current lower limit of $92.5$ GeV~\cite{LEPCH}. Various
strategies for using the $W W^*$ or the $b \bar b$ decay mode for 
$h$ to enhance the mass reach of the Tevatron have been suggested~\cite{tao}.

At LHC energies, due to the large available gluon fluxes and the 
large value of \mt, Eq.~\ref{e10} is the dominant production process 
for all values of \mh\ upto the upper bounds discussed in
section~\ref{twoone}. The total production cross-section goes from
$\sim 20$ pb to $\sim 0.1$ pb as \mh\ goes from 100 to 1000 GeV.
For a superheavy Higgs ($\gsim$ 800 GeV i.e. above the
highest upper bound of sec.~\ref{twoone}), the process of Eq.~\ref{e13} 
has significant cross-section. The higher order QCD corrections
to the gluon induced higgs production are significant and have been 
included in the available theoretical predictions shown in 
\begin{figure}[htb]
\begin{center}
\mbox{\epsfig{file=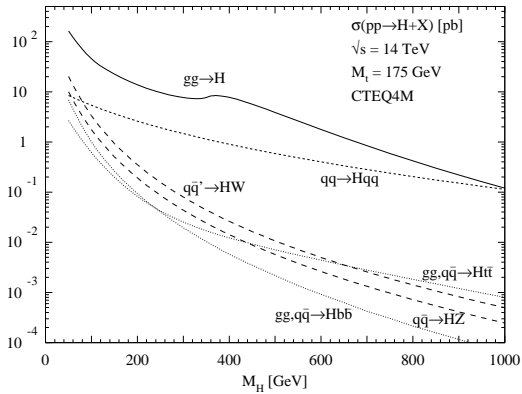,height=60mm}}
\vspace{-0.5cm}
\end{center}
\caption{Production cross-section for the SM Higgs at LHC \label{figure5}}
\end{figure}
Fig.~\ref{figure5}~\cite{spira}.
These predictions, at the LHC, have a typical theoretical uncertainty of 
$\sim 20\%$ due to the parton densities. 

Since LEP-2 has already ruled out, from direct search, $m_h < 92.5$ 
GeV~\cite{LEPCH}, the range of masses of interest to LHC divides neatly into
two parts : $92.5 < \mh \lsim 140$ GeV and $\mh\ > 140$ GeV. In the 
first mass region the dominant decay mode of the $h$ is into a $b \bar b$ final 
state which has a QCD background about $10^3$ higher than the signal. Hence,
in this mass range $h \to \gamgam$ remains the best final state, even with 
a branching ratio $\sim 10^{-3}$.  Even then, the resolution required 
for the $M_{\gamgam}$ measurement has to be $\lsim 1$ GeV $\simeq  .1 \% \; 
\mh$.

The new developements in the detection aspect have been  detector simulations 
for the \gamgam\ and $b \bar b$ mode for the planned detector designs.
\begin{figure}[htb]
\begin{center}
\vspace{-0.8cm}
\mbox{\epsfig{file=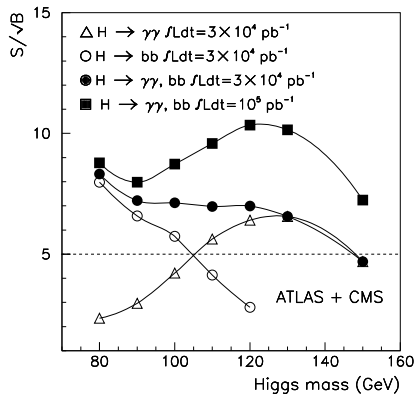,height=60mm}}
\end{center}
\vspace{-.3cm}
\caption{The expected significance level of the SM Higgs signal at LHC for
the intermediate mass region\label{figure6}}
\end{figure}
Fig.~\ref{figure6} taken from the CMS/ATLAS technical 
proposal~\cite{cms1}
shows the  expected $S/\sqrt{B}$ for the SM Higgs in the intermediate mass
range, using the $\gamgam , b \bar b$ modes. The use of $b \bar b$ mode 
for $\mh < 100 $ GeV, is essentially achieved by using the associated 
production of Eq.~\ref{e11}.

Once $\mh\ \gsim 140 $ GeV , the $VV^*$ or $VV$ decay mode is dominant.
As a result, for $140 < \mh\ < 600$ GeV, the 4 lepton final state
following from $h \to ZZ \to l^+l^-l^+l^-$ offers the so called
`gold plated signal' for the Higgs. For $800 < \mh <1000$, the  very
forward jets produced in association with the Higgs in the process 
of Eq.~\ref{e13} provide a much better signature than the `gold plated 
signal'. However, it should be borne in mind, that the unitarity and 
triviality bounds of section~\ref{twoone} imply that in the SM 
$\mh\ \lsim 600-800 $GeV. 

Thus we see that while the detection of an intermediate mass Higgs is difficult 
but feasible at LHC, it will surely require the high luminosity run. 
It should also be kept in mind that  a low value of \mh\ seems to 
be preferred by the precision measurements at LEP and SLC and further
that SUSY also predicts that the lightest scalar in the theory will be in this
mass range. For higher \mh\ values the detection is a certainty at LHC.
However,  to establish such a scalar as {\it the SM} Higgs, one needs to 
establish 
\begin{enumerate}
\item The scalar is CP even and has $J^P = 0^+$,
\item The couplings of the scalar with the fermions and gauge bosons
are proportional to their masses.
\end{enumerate}
This is also essential from the point of view of being able to distinguish 
this scalar from the lightest scalar expected in the MSSM. We see from 
Table~\ref{table1} that the couplings of the scalars to the fermions 
and gauge bosons can be quite different in the MSSM.   

As a matter of fact this issue has been a subject of much investigation
of late~\cite{tevwg,gunion}. The  Snowmass Studies~\cite{tevwg} indicate that 
for a light Higgs ($\mh\ = \mz$) it is possible only to an accuracy of about 
30 \%. It is in this respect that the planned \eplem\ 
colliders~\cite{physrep}
can be a lot of help. At these colliders, the production processes are 
the same as given in Eqs.~\ref{e12}-\ref{e14} where $q (\bar q)$ 
are replaced by $e^- (e^+)$. This is not an appropriate place to give a 
complete discussion of the search prospects for the SM (and MSSM) Higgs
at these colliders. But suffice it to say that if the production is
kinematically allowed, detection of the Higgs at these machines is very simple
as the discovery will be signalled by very striking features of the 
kinematic distributions.  Determination of the spin of the  produced 
particle in this case  will also  be simple as the expected angular
distributions will be very different for even and odd parity. 
Even with this machine one will need a total luminosity of 
$200\;\; {\mathrm fb}^{-1}$, to be able to determine the ratio of 
$BR (h \to c \bar c) / BR(h \to b \bar b)$, to about 
$7 \%$~\cite{gunion}.
The simplest way to determine the CP character of the scalar will be
to produce $h$ in a \gamgam\ collider, which are  being discussed.
There are also interesting invstigations~\cite{gunion} which try to device 
methods to determine the CP character of the scalar using hadron colliders.

For MSSM Higgs the discussion of the actual search possibilities is much more 
involved and has been covered in other talks at this conference~\cite{expthig}.
For the lightest scalar \hzer\  in the  MSSM, the general discussions 
of the intermediate mass Higgs apply, with the proviso that the \gamgam\
branching ratios are smaller for \hzer\ and hence the search that much more 
difficult.  However, since there exist many more scalars in the spectrum
now, one can cover the different regions in the parameter space by looking for 
$A,H$ and $H^\pm$. At low values of \ma\ these other scalars are kinematically
accessible at LHC and also at the NLC. However, even after combining the 
information from various colliders (LEP-II, Tevatron (for the charged higgs
search) and of course LHC), a certain region in the $\ma - \tan \beta$
plane remains inaccessible. This hole can be filled up only after combining
the data from the CMS and ATLAS detector for 3 years of high luminosity run  
of LHC.  Even in this case there exist large region where one will see only the
single light scalar.              

At large \ma\ (which seem to be the values preferred
by the current data on $b \to s \gamma$), the SM higgs and \hzer\ are 
indistinguishable as far as their couplings are concerned, as
can be seen from Table~\ref{table1}. A recent study,
gives the contours of constant values for the ratio 
$$
\frac{BR(c \bar c)/ BR(b \bar b)|_{\hzer}} {BR(c \bar c)/ BR(b \bar b)|_{h}},
$$
as well as a similar ratio for the $WW^*$ and $b \bar b$ widths as a function
of $\tan \beta$ and \ma . 
\begin{figure}[htb]
\begin{center}
\vspace{-0.8cm}
\mbox{\epsfig{file=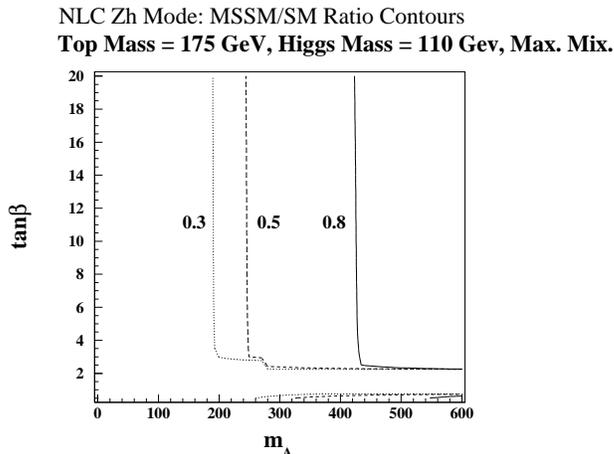,height=70mm}}
\end{center}
\vspace{-0.8cm}
\caption{The ratios of relative branching fractions for the MSSM and SM 
for maximal mixing in squark sector. The specific value of the Higgs mass
used is theoretically disallowed at large \ma\ and around $\tan \beta \sim 2$.
\label{figure7}}
\end{figure}
As we can see from Fig.~\ref{figure7}~\cite{tevwg}, a measurement of this 
ratio to an accuracy of about $10 \%$ will allow distinction 
between the SM Higgs $h$ and MSSM Higgs \hzer\ for $m_A$  as large as 
$\sim 600 $ GeV. As stated above, NLC should therefore be able to do such a job.
Certainly,  the issue of being able to determine the quantum numbers and 
the couplings  of the scalar accurately, forms the subject of a large number of
investigations currently.

\section{Conclusions}
\begin{enumerate}
\item
The experiments at the upcoming hadronic colliders will be able to `discover'
scalar in the entire mass range from 100 GeV to 1000 GeV, with varying degree 
of ease. The lower end is difficult but feasible.
\item 
If the experiments at the Tevatron (Run-II and TeV33, should
it happen) and the LHC do not find a light scalar upto
$\mh\ = 130 (160) $ GeV, it will certainly rule out a class of SUSY 
models; viz. MSSM  and models with minimal extension of the MSSM: the (N)MSSM.
The only way this upper bound can be relaxed is if we enlarge the gauge group 
and introduce additional scales in the problem. 
\item 
If the scalar that is `found' has a mass $\gsim 800$ GeV, it is an indication 
that the EW symmetry breakdown happens via strongly interacting sector.
\item If we {\it do} find a {\bf light} Higgs, then all we can conclude is
that the SM works as an effective theory upto large scales, as discussed 
in section~\ref{twoone}. It will also be consistent with the indirect 
mass limits obtained from precision measurements. This limit
is more or less insensitive to existence of SUSY (or otherwise) due
to large mass scales to which sparticle masses have already been 
pushed by the lack of direct observation of the SUSY particles
\item Since in Supersymmetric models the masses and the decay modes
of the various scalars in the theory are correlated, almost all the region 
in the parameter space of the MSSM can be explored at the experiments
at the LHC~\cite{expthig}.
\item Already with the current data (particularly the 
information on the $b \to s \gamma$) the limits on the MSSM parameter 
space are such that the lightest scalar \hzer\ will be very 
similar in its properties to the SM Higgs $h$. Hence, it is important to 
devise strategies to determine the quantum numbers such as
Spin, Parity, CP etc., of this observed scalar. To that end
a TeV ($\gsim 300 $ GeV) linear \eplem\ collider seems indispensible.
Also, the possibilities of achieving this at the LHC/Run-II/TeV-33 need
to be investigated vigorously.
\end{enumerate}

\section*{Acknowledgements:}
It is a pleasure to thank Profs. Narasimham and Mondal for organising
an excellent conference which provided the forum for many interesting
discussions. This work was in part supported by the Department 
of Science and Technology (India) and the National Science Foundation, 
under NSF-grant-Int-9602567.

\end{document}